\documentclass{ifacconf}

\usepackage{graphicx}      % include this line if your document contains figures
\usepackage{natbib}        % required for bibliography
\usepackage{xcolor}

\usepackage{enumitem}

\usepackage{amsmath}
\usepackage{amssymb}

\usepackage{subcaption}
\usepackage{algorithm}
\usepackage{algpseudocode}
\usepackage{epstopdf}
\usepackage{placeins}
\usepackage{enumitem}

\usepackage{tikz}
\usetikzlibrary{calc}

\allowdisplaybreaks

\newtheorem{theorem}{Theorem}
\newtheorem{assumption}{Assumption}
\newtheorem{lemma}{Lemma}
\newtheorem{proposition}{Proposition}
\newtheorem{remark}{Remark}

\newtheorem{corollaryinner}{Corollary}[proposition] % just the internal version
\newenvironment{corollary}[1][]
 {% #1 is the cross reference label
  \if\relax\detokenize{#1}\relax
  \else
    \ifcsname #1-used\endcsname
      \expandafter\xdef\csname #1-used\endcsname{\the\numexpr\csname #1-used\endcsname+1}%
    \else
      \expandafter\gdef\csname #1-used\endcsname{1}%
    \fi
    \renewcommand{\thecorollaryinner}{\ref{#1}.\csname #1-used\endcsname}%
  \fi
  \corollaryinner
 }
{\endcorollaryinner}

\newcommand{\norm}[1]{\left\lVert#1\right\rVert}

\begin{document}
\begin{frontmatter}

%\title{ Enabling dual control in adaptive MPC methods using exact prediction of parameter sets \thanksref{footnoteinfo}} 
%\title{  Dual adaptive MPC using exact set-membership prediction \thanksref{footnoteinfo}} 
\title{ Dual adaptive MPC using an exact set-membership reformulation\thanksref{footnoteinfo}} 

\thanks[footnoteinfo]{This work was supported by the Swiss National Science Foundation under Grant 200021\_178890.}

\author[First]{Anilkumar Parsi} 
\author[First]{Diyou Liu} 
\author[Third]{Andrea Iannelli}
\author[First]{Roy S. Smith}

\address[First]{Automatic Control Laboratory, ETH Zurich, Zürich 8092 (e-mail: {aparsi,rsmith}@control.ee.ethz.ch, diyliu@student.ethz.ch)}
\address[Third]{University of Stuttgart, Institute for Systems Theory and Automatic Control, Stuttgart 70569 (e-mail: andrea.iannelli@ist.uni-stuttgart.de)}

\begin{abstract}                % Abstract of not more than 250 words.
Adaptive model predictive control (MPC) methods using set-membership identification to reduce parameter uncertainty are considered in this work. %A novel strategy is proposed to model the informativity of future control inputs for the identification task, which is achieved by exactly modeling the equations of set-membership within the MPC optimization problem. 
Strong duality is used to reformulate the set-membership equations exactly within the MPC optimization. A predicted worst-case cost is then used to enable performance-oriented exploration. The proposed approach guarantees robust constraint satisfaction and recursive feasibility. It is shown that method can be implemented using homothetic tube and flexible tube parameterizations of state tubes, and a simulation study demonstrates performance improvement over state-of-the-art controllers. 
\end{abstract}


\end{frontmatter}
%===============================================================================

\section{Introduction} \vspace{-2mm}
In order to optimally control systems affected by model uncertainty, control actions must have two features,
(i) excite the system to obtain new information (called exploration); (ii) use available information to minimize a control cost (called exploitation). Achieving the optimal exploration-exploitation trade-off is the dual control problem, which is a challenging problem in control theory and has been studied for over six decades (\cite{feldbaum1960dual,MESBAH_review}).  With recent advances in computing capabilities, there is an increase in research on controllers which can learn model uncertainties while performing control tasks (e.g., \cite{klenske2016dual,zanon2020safe}), raising the relevance of the dual control problem.

One popular class of controllers which has been studied in this regard is model predictive control (MPC). The ability of MPC to ensure state and input constraint satisfaction in the face of uncertainty has been viewed as an important feature for learning-based control, because it can guarantee safety of the system (\cite{hewing2020learning}). Robust adaptive MPC is one such method, where features from robust MPC methods (such as tube MPC (\cite{kouvaritakis2015model})) are combined with online identification techniques (such as set-membership (\cite{milanese1991})) to ensure safe learning of the model uncertainty. Using various combinations of these two features, multiple robust adaptive MPC algorithms have been proposed in the recent past, e.g., \cite{lorenzen2019,bujarbaruah2020adaptive,SoloPertoAug2020,kohler2021robust}. 

However, most existing robust adaptive MPC algorithms use \textit{passive} exploration. That is, the controller consists of two separately designed components: the MPC optimizer which selects inputs based only on exploitation; and the online identification algorithm which uses the collected measurements to update model uncertainty.
This approach can result in suboptimality, because the dual control effect is unmodeled. One way to partially address this issue is to enforce a perstistency of excitation condition on the control inputs (\cite{lorenzen2019}). However, optimal dual control trade-off would require extensive tuning of these methods, and tuning methods have been proposed in \cite{SoloPertoAug2020}.
%This issue can be partially addressed by enforcing a persistency of excitation condition on the control inputs \cite{lorenzen2019,lu2021robust}, or modifying the objective function to additionally minimize the uncertainty in the parameters \cite{,SoloPertoAug2020}. The disadvantage with such approaches is that extensive tuning would be required to achieve good performance. 
An alternative strategy is to use performance-based exploration, where exploration is induced by modeling the expected performance improvement from reduction in uncertainty. This approach was considered in \cite{parsi2022explicit}, where set-membership identification is used to reduce model uncertainty, and the key novelty consisted of including the effect of identification in the MPC optimizer. Nevertheless, this required approximations which result in overestimating the capabilities of the identification algorithm, and can lead to suboptimal exploratory actions.

In this work, we propose a novel way to exactly reformulate set-membership identification inside MPC optimization problems. The proposed method has two advantages. First, it results in an improved navigation of the dual control trade-off compared to existing methods because the identification is exactly modeled within the MPC controller. Second, the method can be used to generate
dual control algorithms starting from multiple existing formulations of passive adaptive MPC. We highlight this feature by implementing dual control using homothetic and flexible state tube parameterizations, and demonstrate performance improvement using a simulation study.
\vspace{-4mm}
\subsection{Notation} \vspace{-2mm}
The sets of real numbers and non-negative real numbers are denoted by $ \mathbb{R} $ and $ \mathbb{R}_{\ge0} $ respectively. The sequence of integers from $ n_1 $ to $ n_2 $ is represented by $ \mathbb{N}_{n_1}^{n_2} $. 
%For a vector $ b $, $ b^{\intercal} $ represents its transpose, and $ [b]_i $ refers to its $ i ^{th}$ element. 
For a vector $ b $, $||b||_\infty$ represents its $\infty-$norm. The $ i ^{th}$ row of a matrix $ A $ is denoted by $ [A]_{i} $,  and $ \mathbf{1} $ denotes a column vector of appropriate length with all entries 1.  The convex hull operator is represented by $\text{co}\{\cdot\}$. The notation $ a_{l|k} $ denotes the value of $ a $ at time $ k{+}l $ computed at time $ k $, $I_n$ denotes the identity matrix with $n$ rows, and $\oplus$ denotes the Minkowski sum of two sets.
\vspace{-4mm}
\section{Problem formulation} \label{Sec:ProbForm}   \vspace{-2mm}
Consider an uncertain, linear system described by
\begin{equation}\label{eq:Dynamics}
    x_{t+1} = A(\theta) x_t + B(\theta) u_t + w_t,
\end{equation}
where  $x_t {\in} \mathbb{R}^n$ represents the state, $u_t {\in} \mathbb{R}^m$ the control input and $w_t {\in} \mathbb{R}^n$ the additive disturbance at time $t$. The state space matrices have the parametric description
\begin{align} \label{eq:Parameterization}
\begin{split}
    A(\theta) = A_0 + \displaystyle\sum_{i=1}^{p} A_i [\theta]_i, \quad  B(\theta) = B_0 + \displaystyle\sum_{i=1}^{p} B_i [\theta]_i,
\end{split}
\end{align}
where $\theta {\in} \mathbb{R}^p$ is a constant but unknown parameter, and $\{A_i,B_i\}_{i=0}^{p}$ are known matrices which model structured uncertainty. The true value of $\theta=\theta^*$ is known to lie inside a bounded set
\begin{equation}\label{eq:ParameterBounds}
    \Theta := \{\theta \in \mathbb{R}^p | H_{\theta} \theta \le h_{\theta} \},
\end{equation}
where $H_{\theta} \in\mathbb{R}^{n_\theta \times p}$ and $h_{\theta} \in \mathbb{R}^{n_{\theta}}$. The disturbance $w_t$ always lies within the bounded set
\begin{equation}\label{eq:DisturbanceBounds}
    \mathbb{W}:= \{w\in\mathbb{R}^n |  H_{w} w \le h_{w} \},
\end{equation}
where $ H_w \in \mathbb{R}^{n_w\times n}, h_w\in\mathbb{R}^{n_w}$.  The state of the system is perfectly measured at each time step. Given an initial state $x_0$, the objective is to track a reference trajectory given by  $r_t\in\mathbb{R}^{n}$ for time steps t in $ [0,T]$. This objective is specified as a minimization of the worst-case deviation of the states and inputs from the optimal reference, i.e.,
\begin{equation}\label{eq:Cost_to_min}
    J= \min_{u_t} \max_{w_t \in \mathbb{W}} \sum_{t=0}^{T} \norm{Q(x_t{-}r_{t})}_{\infty} {+} \norm{R(u_t{-}u^*_t)}_{\infty},
\end{equation}
where $Q$ and $R$ are positive definite matrices, and $\{u^*_t\}_{t=0}^{T}$ is a sequence of (possibly) unknown input setpoints which track the reference trajectory while satisfying the dynamics of the true system. The states and inputs of the system are constrained to lie inside the compact set
\begin{equation} \label{eq:Constraints}
\mathbb{Z} = \left\{(x,u) \in \mathbb{R}^n \times \mathbb{R}^m \bigr|  F x + G u \le \mathbf{1}\right\} ,
\end{equation}
where $ F \in \mathbb{R}^{n_c \times n} $ and  $ G \in \mathbb{R}^{n_c \times m} $ are known matrices. 

In the following sections, a dual control algorithm will be described to perform the above task. In Section \ref{Sec:RAMPC}, an online identification scheme will be used to reduce uncertainty in $\theta$, and a robust state tube will be constructed to ensure constraint satisfaction. In Section \ref{Sec:PST_Exploration}, future set-membership is formulated as a function of the MPC input variables, and the MPC cost function is defined using a predicted state tube to enable performance-based exploration. The dual adaptive MPC algorithm and its properties are discussed in Section \ref{Sec:DualAMPC}. The proposed method will be implemented using two popular parameterizations of state tubes from literature, namely homothetic tubes (HT)  (\cite{lorenzen2019}) and flexible tubes (FT)  (\cite{lu2019robust}).

\section{Parameter identification and robust state tubes}\label{Sec:RAMPC} \vspace{-2mm}
\subsection{Set membership and parameter estimates} \vspace{-2mm}
In order to reduce the uncertainty in $\theta$, set-membership identification (\cite{milanese1991}) is used to construct a sequence of sets $\{\Theta_k\}_{k=0}^{T}$ satisfying 
\begin{align}\label{eq:SetInclusion}
\begin{split}
    \Theta_{k-1} &\cap \Delta_k \subseteq \Theta_k \subseteq \Theta_{k-1}, \quad  \Theta_0 = \Theta, \\
    \Delta_k = \{&\theta | x_{i+1} {-} A(\theta)x_{i} {-} B(\theta)u_{i} \in \mathbb{W} , \: \forall i \in \mathbb{N}_{k-\tau}^{k-1} \} \\
            = \{ &\theta | -H_w D_i\theta \leq h_w + H_w d_{i+1},\:  \forall i \in \mathbb{N}_{k-\tau}^{k-1} \},
\end{split}
\end{align}
where $\Delta_k$ is the set of all parameters $\theta$ that could have generated the last $\tau$ measurements $\{ x_i\}_{i=k-\tau+1}^{k}$, and % $D_{i-1} \in\mathbb{R}^{n\times p}$ and $d_{i}\in \mathbb{R}^{n}$ are 
\begin{align}
\begin{split}
    D_i &:= D(x_{i},u_{i}) = \begin{bmatrix}
         A_1 x_i {+} B_1 u_i, & \ldots, & A_p x_i {+} B_p u_i
    \end{bmatrix},  \\
    d_{i+1} &:= \: A_0 x_k + B_0 u_k - x_{i+1}, \quad \forall i \in \mathbb{N}_{k-\tau}^{k-1}.
\end{split}
\end{align} 

The set $\Theta_k$ is described by a fixed number of hyperplanes for computational efficiency, i.e., 
\begin{equation}\label{eq:Theta_k}
    \Theta_k := \{\theta| H_{\theta} \theta \le h_{\theta_k} \}.
\end{equation} 
Here $h_{\theta_k}$ is initialized with the initial estimate of uncertainty $h_{\theta}$, and is updated such that \eqref{eq:SetInclusion} is satisfied by solving a set of linear programs, as proposed in \cite{lorenzen2019}. Whereas HT adaptive MPC methods allow $H_{\theta}$ in \eqref{eq:Theta_k} to be arbitrarily chosen, FT methods further restrict $\Theta_k$ to polytopes with hyperplanes parallel to coordinate axes (or hyperboxes). The numerical advantage is that, the vertices of $\Theta_k$, denoted by $\{\theta^j_k\}_{j=1}^{q_\theta} $, can be readily computed as a predefined linear combination of $h_{\theta_k}$ (\cite{diyou2022}). 

\begin{lemma}\cite[Theorem 14(ii)]{lorenzen2019}
    Under the set-membership scheme \eqref{eq:SetInclusion}-\eqref{eq:Theta_k}, $\theta^*\in \Theta_k $ for all $ k\ge 0.$ %the true parameter $\theta^*$ always lies inside $\Theta_k$.
\end{lemma}
\vspace{-2mm}

Additionally, an estimate $\bar{\theta}_k \in \Theta_k $ of the parameter $\theta^*$ is computed  using a least mean squares filter
%in order to define the input setpoint to be tracked. One approach is to use a least mean squares filter, with the update equations
\begin{align}\label{eq:LMS}
\begin{split}
    & \tilde{\theta}_{k+1} = \Bar{\theta}_{k} {+} \mu D(x_{k}, u_{k})^T(x_{k+1} {-} A(\Bar{\theta}_k)x_k {-} B(\Bar{\theta}_k)u_k) ,
    \\
    & \Bar{\theta}_{k+1} = \Pi_{\Theta_{k}}(\tilde{\theta}_{k+1}),
\end{split}
\end{align}
where $\Pi_{\Theta_{k}}$ represents a projection operator, and $\bar{\theta}_0$ is an initial guess.

\vspace{-3mm}
\subsection{Control policies and tracking formulation} \vspace{-2mm}
The control input is parameterized as
\begin{equation}\label{eq:InputParameterization}
    u_{l|k}(x) = K({x}-r_{k+l}) + v_{l|k}, \quad \forall l\in \mathbb{N}_{0}^{N-1},
\end{equation}
where $v_{l|k}$ are online optimization variables, $K$ is a stabilizing feedback gain and $N$ is the MPC prediction horizon. 
\begin{assumption}\label{As:Feedback}
	The parameter set $\Theta$ is such that there exists a feedback gain $K$ which asymptotically stabilizes $ A_{\text{cl}}(\theta)  = A( \theta) + B(\theta)K, \: \forall \theta \in \Theta$.
\end{assumption}
\vspace{-2mm}

The input setpoints to be tracked are unknown due to the uncertainty in the true parameter. Hence, the MPC problem estimates $u^*_{k+l}$  using $\bar{\theta}_k$ and optimization variables $\bar{u}_{l|k}$ such that
\begin{align} \label{eq:SetpointSubspace}
\begin{split}
    \begin{bmatrix}
    A(\bar{\theta}_k) & B(\bar{\theta}_k)
    \end{bmatrix} \begin{bmatrix}
    r_{k+l}\\\bar{u}_{l|k}
    \end{bmatrix} & = \begin{bmatrix}
    r_{k+l+1}
    \end{bmatrix} , \quad \forall l \in \mathbb{N}_{0}^{N-1},\\
    \begin{bmatrix}
    A(\bar{\theta}_k) & B(\bar{\theta}_k)
    \end{bmatrix} \begin{bmatrix}
    r_{k+N}\\\bar{u}_{N|k}
    \end{bmatrix} & = \begin{bmatrix}
    r_{k+N}
\end{bmatrix}.
\end{split}
\end{align} 
The setpoint for timestep $k+N$ is chosen to be an equilibrium point, in order to define the terminal components of MPC (discussed in Section \ref{Sec:TerminalSets}). Under assumptions on the rank of the system matrices and the number of available inputs \cite[Assumption~2]{parsi2022explicit}, it can be guaranteed that there always exists a feasible input sequence satisfying \eqref{eq:SetpointSubspace}.
\vspace{-3mm}
\subsection{State tube construction} \vspace{-2mm}
Using the control policy \eqref{eq:InputParameterization}, a sequence of sets $\{\mathbb{X}_{l|k}\}_{l=0}^{N}$ is constructed such that $\mathbb{X}_{0|k} \ni x_k$ and for all $x\in \mathbb{X}_{l|k}, \theta \in \Theta_k$ and $ w\in \mathbb{W}$,
\begin{align}\label{eq:SetDynamics}
    \mathbb{X}_{l+1|k} \ni A(\theta)x + B(\theta)u(x) + w, \quad l\in \mathbb{N}_{0}^{N-1}.
\end{align}
This sequence of sets is called the robust state tube (RST), as it contains all feasible state trajectories within the MPC prediction horizon. The RST is constructed by outer approximating the reachable states using parameterized sets for computational efficiency.

\vspace{-3mm}
\subsubsection{Homothetic tubes.}
In an HT formulation, the RST is parameterized as
\begin{align}\label{eq:HT_Parameterization}
\mathbb{X}^H_{l|k} &= \{z_{l|k}\} \oplus \alpha_{l|k} \mathbb{X}_{0},
\end{align}
where $z_{l|k}\in \mathbb{R}^{n}$ and $\alpha_{l|k}\in \mathbb{R}_{\ge 0}$ are translation and scaling variables and $\mathbb{X}_0$ is a predefined polytope described in both hyperplane and vertex forms 
\begin{align}
 \mathbb{X}^H_0 := \{x| H_x x \le \mathbf{1}\} = \text{co}\{x^{1},x^{2},\ldots,x^{q}\}.
\end{align}
The following notation is introduced
\begin{align}\label{eq:x_jlk}
\begin{split}
x_{l|k}^{j} &= z_{l|k} + \alpha_{l|k} x^{j}, \quad u_{l|k}^{j} {=} K(x_{l|k}^{j} - r_{k+l}) {+} v_{l|k},  \\
 D_{l|k}^{j} &= D(x_{l|k}^{j},u_{l|k}^{j}), \quad  d_{l|k}^{j} {=} A_0 x_{l|k}^{j} {+} B_0 u_{l|k}^{j} - z_{l+1|k}, \\
 [\bar{f}]_{i} &= \underset{x\in \mathbb{X}_0}{\text{max}} [F+GK]_i x,\quad [\bar{w}]_{j} = \underset{w\in \mathbb{W}}{\text{max}}  \: [H_x]_j w,
\end{split}
\end{align}
where $\Bar{f}$ and $\Bar{w}$ are computed offline. 
 
\begin{proposition}\cite[Proposition 9]{lorenzen2019}\label{Pr:HT_RST}
	Let the RST  be parameterized according to  \eqref{eq:HT_Parameterization}. Then, \eqref{eq:Constraints} and \eqref{eq:SetDynamics} are satisfied if,  $\forall \: j{\in} \mathbb{N}_{1}^{q} $ and $ l{\in} \mathbb{N}_{0}^{N-1}$, there exist $ \Lambda_{l|k}^{H,j} \in \mathbb{R}^{n_x\times n_\theta}_{\ge0}$  such that 
	\begin{subequations}\label{eq:HT_RST}
		\begin{align}
		(F+GK)z_{l|k} + G (v_{l|k} - Kr_{k+l}) + \alpha_{l|k}\bar{f} &\le \mathbf{1},\\
		-H_x z_{0|k} -\alpha_{0|k}\mathbf{1} &\le -H_x x_k ,		\label{eq:x0Constraint}\\
		\Lambda_{l|k}^{H,j} h_{\theta_k} + H_x d_{l|k}^{j} -\alpha_{l+1|k} \mathbf{1} &\le -\bar{w} ,\label{eq:InclusionIneq}\\
		H_x D_{l|k}^{j} &= \Lambda_{l|k}^{H,j} H_{\theta} \label{eq:InclusionEqual}.
		\end{align} 
	\end{subequations}
\end{proposition}
The reformulation \eqref{eq:InclusionIneq}-\eqref{eq:InclusionEqual} is obtained by imposing \eqref{eq:SetDynamics} at each vertex of $\mathbb{X}_{l|k}$, and using strong duality (\cite{boyd2004convex}) to ensure \eqref{eq:SetDynamics} for all $\theta \in \Theta_k$. 

\vspace{-3mm}
\subsubsection{Flexible tubes. }
In an FT formulation, the state tube is parameterized as
\begin{align}\label{eq:FT_Parameterization}
\mathbb{X}^F_{l|k} &= \{x| H_x (x-r_{k+l}) \leq  h_{x_{l|k}}\}, \qquad \forall l \in {\mathbb{N}}_0^{N},
\end{align}
where $H_x$ is a matrix chosen offline, and $h_{x_{l|k}}$ are online optimization variables. The following proposition can then be used to reformulate the tube propagation constraints.
\begin{proposition}{\cite[Lemmas 8 and 9]{lu2019robust}}\label{Pr:FT_RST}
	Let the RST be parameterized according to \eqref{eq:FT_Parameterization}. Then, \eqref{eq:Constraints} and \eqref{eq:SetDynamics} are satisfied if, $\forall  \: i {\in} \mathbb{N}_{1}^{n_x}$, $ j {\in} \mathbb{N}_{1}^{q_{\theta}}$ and $ l{\in} \mathbb{N}_{0}^{N-1}$,
\begin{align}\label{eq:FT_RST}
    \begin{split}
      \Lambda_{1,l|k}^F(h_{x_{l|k}} + H_x r_{k+l}) + G(v_{l|k} - K r_{k+l}) &\leq \mathbf{1},\\
        H_x(x_k - r_k) &\leq h_{x_{0|k}}, \\
        [1 \: \: {\theta_k^j}^T] \Lambda_{2,l|k}^{F,i} (h_{x_{l|k}} + H_x r_{k+l})  - [H_x]_i r_{k+l+1}& -    \\
        [H_x]_i(B(\theta_k^j)(Kr_{k+l}-v_{l|k})) + [\Bar{w}]_i \leq [&h_{x_{l+1|k}}]_i , 
    \end{split}		
\end{align} 
where $\Lambda_{1,l|k}^F {\in} \mathbb{R}_{\geq 0}^{n_c \times n_x} $ and $\Lambda_{2,l|k}^{F,i} {\in} \mathbb{R}^{(p+1) \times n_x}$ are Lagrange multipliers and are computed offline by solving $\forall l \in \mathbb{N}_{0}^{N-1}$,
\begin{align}\label{eq:Flexible_Constraints_Lambda}
\begin{split}
   \forall s \in \mathbb{N}_1^{n_c},\quad  [\Lambda_{1,l|k}^F]_s = & \arg \min\limits_{\lambda } \lambda (\mathbf{1} + \mu H_x r_{k+l})\\
    \text{s.t.} \qquad \lambda &\geq 0, \quad \lambda H_x = [F+GK]_s,
\end{split}
\end{align}
\begin{align}\label{eq:Flexible_Propagation_Lambda}
\begin{split}    
        \Lambda_{2,l|k}^{F,i} = \arg &\min\limits_{\Lambda}  \max\limits_{j \in \mathbb{N}_1^{q_\theta}} [1 \: \: {\theta_k^j}^T] \Lambda (\mathbf{1} + \mu H_x r_{k+l})\\
        \text{s.t.} \quad [1 &\quad {\theta_k^j}^T]\Lambda \geq 0, \quad \forall j \in \mathbb{N}_1^{n_\theta},\\
        \Lambda H_x = &\begin{bmatrix}
          [H_x]_i (A_0 {+} B_0 K)^T &
          \ldots &
          [H_x]_i (A_p {+} B_p K)^T  \end{bmatrix}^T,
\end{split}
\end{align}
where  $\mu\in \mathbb{R}_{\ge 0}$.
\end{proposition}

The reformulation \eqref{eq:FT_RST} is obtained by imposing \eqref{eq:SetDynamics} at each vertex of $\Theta_k$, and using duality to ensure \eqref{eq:SetDynamics} for all $x\in\mathbb{X}_{l|k}$. In addition, duality is also used to reformulate \eqref{eq:Constraints}. The tuning factor $\mu$  facilitates the reduction of conservatism by weighing the effects of both $h_{x_{l|k}}$ and $r_{k+l}$ on \eqref{eq:Constraints} and \eqref{eq:SetDynamics} while choosing the Lagrange multipliers offline.  Note that Proposition \ref{Pr:FT_RST} is a slight modification of the result in \cite{lu2019robust}, where the reference was the origin. 
%The factor $\mu$ is used because the offline choice of $\Lambda_{1,l|k}^F$ and $\Lambda_{2,l|k}^{F,i} $ affects the conservatism of the FT's outer approximation of \eqref{eq:SetDynamics} in \eqref{eq:FT_RST}. 

\begin{remark}
The FT formulation requires that both hyperplanes and vertices of $\Theta_k$ are available, and HT requires the same for $\mathbb{X}_{0}$. Moreover, the FT formulation computes the Lagrange multipliers offline for faster online solve times, which results in conservatism (\cite{kohler2019linear}).
\end{remark}

\subsection{Terminal Sets} \label{Sec:TerminalSets}  \vspace{-2mm}
In order to ensure recursive feasibility, terminal sets are used. That is, the last set in the RST must be an invariant set under a terminal control law of the form
\begin{equation}\label{eq:TerminalControlLaw}
    u_{N|k}(x) = K(x-\tilde{x}_{N|k}) + v_{N|k},
\end{equation}
where $v_{N|k}$ are MPC optimization variables and $\tilde{x}_{N|k}$ is chosen as specified later. Then, the invariance conditions can be written as, 
for all $x\in\mathbb{X}_{N|k}$, $\theta\in \Theta_k$ and $w \in \mathbb{W}$,
\begin{equation}\label{eq:TerminalCondition}
    \mathbb{X}_{N|k} \ni A(\theta)x {+} B(\theta)u_{N|k}(x) {+} w,  \: \: (x, u_{N|k}(x)) \in \mathbb{Z}.
\end{equation} 

\begin{proposition}\label{Prop:TerminalConditions}
 The constraints \eqref{eq:TerminalCondition}  are satisfied
 \begin{enumerate}[label=\alph*)]
     \item under the HT parameterization \eqref{eq:HT_Parameterization} and $\tilde{x}_{N|k} {=} z_{N|k}$, if $\exists \Lambda_{N|k}^{H,j} \in \mathbb{R}^{n_x\times n_\theta}_{\ge 0}$ such that $\forall j \in \mathbb{N}_{1}^{q}$
        \begin{align}\label{eq:HT_TermCond}
            \begin{split}
                F z_{N|k} + Gv_{N|k} + \alpha_{N|k}\bar{f} &\le \mathbf{1},\\
                \Lambda_{N|k}^{H,j} h_{\theta_k} + H_x d_{N|k}^{j} -\alpha_{N|k} \mathbf{1} &\le -\bar{w}, \\
                \Lambda_{N|k}^{H,j}\ge 0, \quad  H_x D_{N|k}^{j} &= \Lambda_{N|k}^{H,j} H_{\theta}.
            \end{split}
        \end{align} 
        
        \item under the FT parameterization \eqref{eq:FT_Parameterization} and $\tilde{x}_{N|k} {=} r_{k{+}N}$, if  $\forall \: i {\in} \mathbb{N}_{1}^{n_x}$, $ j {\in} \mathbb{N}_{1}^{q_{\theta}}$,
        \begin{align}\label{eq:FT_TermCond}
            \begin{split}
            \Lambda_{1,N|k}^F(h_{x_{N|k}} {+} H_x r_{k+N}) + G(v_{N|k} - K r_{k+N}) \leq &\mathbf{1},\\
            [1 \: \: {\theta_k^j}^T] \Lambda_{2,N|k}^{F,i} (h_{x_{N|k}} + H_x r_{k+N}) -  [H_x]_i r_{k+N} - \: \:& \\
            [H_x]_i(B(\theta_k^j)(K r_{k+N} -v_{N|k})) + \Bar{w} \leq [h_{x_{N|k}}&]_i.
            \end{split} 
        \end{align}
 \end{enumerate}
\end{proposition} 
The proof  is similar to that of Propositions \ref{Pr:HT_RST} and \ref{Pr:FT_RST}, and is omitted. A detailed proof can be found in \cite{diyou2022}. 

\begin{assumption}\label{As:Hx}
    In both HT and FT methods, the predefined state tube matrix $H_x$ is chosen such that for all $\theta\in \Theta$ and $x\in \{x|H_x x \le \mathbf{1}\}$, $\exists \lambda_c \in [0,1)$ satisfying
    \begin{equation}
        H_x \bigr( A(\theta)x + B(\theta)Kx\bigr) \le \lambda_c \mathbf{1}.
    \end{equation}
\end{assumption}
Assumption \ref{As:Hx} imposes contractivity on the state tube shape when $w_k$ is zero. Such an assumption is common in literature (\cite{lu2019robust}). % and can be used to derive necessary conditions for the feasibility of terminal constraints \cite[Proposition 2]{parsi2022explicit}. 
In this work, $\lambda_c$ will be used to define the terminal cost in Section \ref{Sec:CostFunction}.

\vspace{-2mm}
\section{Predicted state tube for Exploration }\label{Sec:PST_Exploration}  \vspace{-2mm}
In this section, set-membership identification is formulated as a function of MPC control input variables. Then, a predicted state tube (PST) is constructed to contain state trajectories as a function of predicted uncertainty sets. Finally, the MPC cost function is defined such that performance-based exploration is induced.

\vspace{-2mm}
\subsection{Predicted set-membership identification} \vspace{-2mm}
In order to predict the effect of the future control inputs on identification, a sequence of predicted measurements $\{\hat{x}_{l|k}\}_{l=0}^{N_p}$ is defined such that
\begin{align}\label{eq:xhat}
\begin{split}
\hat{x}_{0|k} &= x_k, \quad \hat{x}_{l+1|k} = A(\bar{\theta}_k) \hat{x}_{l|k} + B(\bar{\theta}_k) \hat{u}_{l|k} ,\\
\hat{u}_{l|k} &= K (\hat{x}_{l|k} - r_{k+l}) + v_{l|k}, \quad \forall l\in\mathbb{N}_{0}^{N_{p} -1} ,
\end{split}
\end{align}
where $N_p{\in} \mathbb{N}_{2}^{N}$ is the user-defined look-ahead horizon. A large $N_p$ improves the modeling of the dual effect inside MPC, but also increases the computational complexity.  A prediction of $\Delta_{k+l}$ can then be constructed using \eqref{eq:xhat} as 
\begin{align}\label{eq:Delta_hat}
\hat{\Delta}_{ l|k} &:= \{\theta \: | \hat{x}_{i+1|k} {-} A(\theta) \hat{x}_{i|k} {-} B(\theta) \hat{u}_{i|k} \in \mathbb{W} \: \forall i \in \mathbb{N}_{l-\tau}^{l-1}  \} \nonumber\\
& = \{\theta \: | \hat{H}_{\Delta_{l|k}} \theta \leq \hat{h}_{\Delta_{l|k}}\},
\end{align}
where $\hat{x}_{i|k}{=}x_{k+i}, \hat{u}_{i|k}{=}u_{k+i} $ when $ i{<} 0$. Note that $\hat{H}_{\Delta_{l|k}}$ and $ \hat{h}_{\Delta_{l|k}}$ depend on future control inputs, but the dependence is dropped for clarity of presentation. Using \eqref{eq:Delta_hat}, the predicted parameter sets can be defined as
\begin{align}\label{eq:Theta_hat}
    \hat{\Theta}_{0|k} &:= \Theta_k, \: \hat{\Theta}_{l|k} := \{\theta | H_\theta \theta \leq \hat{h}_{\theta_{l|k}}\}, \quad l \in \mathbb{N}_{0}^{N}.
\end{align}
In \eqref{eq:Theta_hat}, $\hat{h}_{\theta_{0|k}} {=} h_{\theta_k}$, and for $l\in\mathbb{N}_{1}^{N_p}$, $\hat{h}_{\theta_{l|k}}$ is computed as
\begin{align}\label{eq:EPPS_RHS}
\begin{split}
    \forall i \in \mathbb{N}_{1}^{n_{\theta}}, \: [\hat{h}_{\theta_{l|k}}]_i &= \mathop{\max}\limits_{\theta} \:  [H_{\theta}]_{i}\theta\\
    \text{s.t.} \quad    H_\theta \theta &\le \hat{h}_{\theta_{l-1|k}} , \quad \hat{H}_{\Delta_{l|k}} \theta \le \hat{h}_{\Delta_{l|k}}.
\end{split}
\end{align}
For $l{\in}\mathbb{N}_{N_p + 1}^{N}$,   $\hat{h}_{\theta_{l|k}} {=} \hat{h}_{\theta_{N_p|k}}$, i.e., no further identification is modeled. Because $\hat{\Delta}_{l|k}$ depends on future control inputs, $\hat{h}_{\theta_{l|k}}$ cannot be explicitly computed before solving the optimization problem. Instead, the following dual reformulation is used, 
\begin{align} \label{eq:EPPS_RHS_Dual}
\begin{split}
    \hat{h}_{\theta_{l|k}} =  \hat{\Psi}_{l|k} \begin{bmatrix}
       \hat{h}_{\theta_{l-1|k}}\\ \hat{h}_{\Delta_{l|k}}
        \end{bmatrix}, \quad 
  \hat{\Psi}_{l|k} \begin{bmatrix}
        H_\theta \\ \hat{H}_{\Delta_{l|k}}
        \end{bmatrix}  = H_{\theta},  \:  \hat{\Psi}_{l|k}  \geq 0,
\end{split}
\end{align}
where $\hat{\Psi}_{l|k}$ represent Lagrange multipliers. Using strong duality, 
it can be seen that there exist $\hat{\Psi}_{l|k}$ such that \eqref{eq:EPPS_RHS_Dual} holds. Thus, using $\hat{\Psi}_{l|k}$ as MPC optimization variables, \eqref{eq:EPPS_RHS} can be replaced by \eqref{eq:EPPS_RHS_Dual}.

\begin{remark}[Bilinearities]
    The constraints \eqref{eq:EPPS_RHS_Dual} include the product of $\hat{\Psi}_{l|k}$ with other optimization variables. Thus, \eqref{eq:EPPS_RHS_Dual} results in bilinear constraints, and the MPC optimization problem is nonconvex.
\end{remark}\vspace{-3mm}

A similar dual adaptive MPC method has been proposed for HT in \cite{parsi2022explicit}, but it approximates set-membership online using an intersection of ${\Theta}_{k}$ and $\hat{\Delta}_{l|k}$. This results in MPC overestimating the capabilities of the actual set-membership identification. %Moreover, the approximation cannot be applied to FT, because the vertices of $\hat{\Theta}_{l|k}$ cannot be computed online.

\subsection{Predicted state tubes} \vspace{-2mm}
The effect of the future identification on the state trajectories can be estimated by defining a predicted state tube $\{ \hat{\mathbb{X}}_{l|k} \}_{l=0}^{N}$ such that  $x_k \in \hat{\mathbb{X}}_{0|k}$ and $\forall l{\in }\mathbb{N}_{0}^{N-1}$
%In order to estimate the effect of future identification on performance, the predicted parameter sets are used to construct a PST $\{ \hat{\mathbb{X}}_{l|k} \}_{l=0}^{N}$ such that  $x_k \in \hat{\mathbb{X}}_{0|k}$ and
\begin{align}\label{eq:PST_dynamics}
    \hat{\mathbb{X}}_{l{+}1|k} {\ni} A(\theta)x {+} B(\theta)u(x) {+} w,\:  \forall x{\in} \hat{\mathbb{X}}_{l|k}, \theta{\in} \hat{\Theta}_{l|k}, \forall w {\in} \mathbb{W}.
\end{align} 
%Note the similarity to the construction of the RST in \eqref{eq:SetDynamics}. In fact, it always holds that $\hat{\mathbb{X}}_{l|k} \subseteq \mathbb{X}_{l|k}$, because $\hat{\Theta}_{l|k} \subseteq \Theta_{k}$. 
It can be seen from \eqref{eq:SetDynamics} that $\forall \: l\in \mathbb{N}_{0}^{N-1}, $ $ \hat{\mathbb{X}}_{l|k} {\subseteq }\mathbb{X}_{l|k}$ because $\hat{\Theta}_{l|k} \subseteq \Theta_{k}$ by construction \eqref{eq:EPPS_RHS}. 

\subsubsection{Homothetic tubes. }
In the HT parameterization, the PST is represented by polytopes of the form
\begin{align}\label{eq:HT_PST_Parameterization}
\hat{\mathbb{X}}^H_{l|k} &= \{\hat{z}_{l|k}\} \oplus \hat{\alpha}_{l|k} \mathbb{X}_{0},  \quad \forall l \in {\mathbb{N}}_0^{N},
\end{align}
where $\hat{z}_{l|k}\in \mathbb{R}^{n_x}$ and $\hat{\alpha}_{l|k}\in \mathbb{R}_{\ge 0}$ are translation and scaling variables. Then, the following is a direct extension of Proposition \ref{Pr:HT_RST}.

\begin{corollary}[Pr:HT_RST]
The PST dynamics \eqref{eq:PST_dynamics} are satisfied under the HT parameterization \eqref{eq:HT_PST_Parameterization}  if  $\forall  j{\in} \mathbb{N}_{1}^{q} $, $ l{\in} \mathbb{N}_{0}^{N-1}$, there exist $ \hat{\Lambda}_{l|k}^{H,j} \in \mathbb{R}^{n_x\times n_\theta}_{\ge0}$  such that 
    \begin{align}\label{eq:HT_PST_propagation}
        \begin{split}
        -H_x \hat{z}_{0|k} -\hat{\alpha}_{0|k}\mathbf{1} \le -H_x x_k ,	\quad  H_x \hat{D}_{l|k}^{j} &= \hat{\Lambda}_{l|k}^{H,j} H_{\theta},	\\
        \hat{\Lambda}_{l|k}^{H,j} \hat{h}_{\theta_{l|k}} + H_x \hat{d}_{l|k}^{j} -\hat{\alpha}_{l+1|k} \mathbf{1} &\le -\bar{w} ,
        \end{split} 
    \end{align}
    where $\hat{D}_{l|k}^{j}$ and $\hat{d}_{l|k}^{j}$ are predictions of ${D}_{l|k}^{j}$ and ${d}_{l|k}^{j}$ defined in \eqref{eq:x_jlk}.
\end{corollary}

\subsubsection{Flexible tubes. }
In the FT parameterization the, PST is represented by polytopes of the form
\begin{align}\label{eq:FT_PST_Parameterization}
\hat{\mathbb{X}}^F_{l|k} &= \{x| H_x (x-r_{k+l}) \leq  \hat{h}_{x_{l|k}}\}, \quad \forall l \in {\mathbb{N}}_0^{N},
\end{align}
where $\hat{h}_{x_{l|k}}\in \mathbb{R}^{n_x}$ are optimization variables. Then, the following is a direct extension of Proposition \ref{Pr:FT_RST}.
\begin{corollary}[Pr:FT_RST]
The PST dynamics \eqref{eq:PST_dynamics} are satisfied under the FT parameterization \eqref{eq:FT_PST_Parameterization} if $\forall i \in \mathbb{N}_{1}^{n_x}$, $ j \in \mathbb{N}_{1}^{q_{\theta}}$ and $ l{\in} \mathbb{N}_{0}^{N-1}$ 
\begin{align}\label{eq:FT_PST_propagation}
    \begin{split}
        H_x(x_k - r_k) &\leq \hat{h}_{x_{0|k}}, \\
        [1 \: \: \hat{\theta}_{l|k}^{j^T}] \Lambda_{2,l|k}^{F,i} (\hat{h}_{x_{l|k}} + H_x r_{k+l})  - [H_x&]_i r_{k+l+1} -    \\
        [H_x]_i(B(\hat{\theta}_{l|k}^j)(Kr_{k+l}-v_{l|k})) + \Bar{w} &\leq [\hat{h}_{x_{l+1|k}}]_i , 
    \end{split}		
\end{align} 
where $\Lambda_{2,l|k}^{F,i}$ are computed offline as defined in \eqref{eq:Flexible_Propagation_Lambda}. 
\end{corollary}
In \eqref{eq:FT_PST_propagation}, $\hat{\theta}_{l|k}^j$  can be formulated as linear combination of terms in $\hat{h}_{\theta_{l|k}}$  because $\hat{\Theta}_{l|k}$ is  restricted to hyperboxes. % Moreover, \eqref{eq:FT_PST_propagation} also results in bilinear constraints, due to the product of $\hat{\theta}_{l|k}^j$ and $\hat{h}_{x_{l|k}}$. 

\subsection{MPC Cost function} \label{Sec:CostFunction} \vspace{-2mm}
%The MPC cost function is now defined using the PST such that performance-based exploration is induced. 
By design, the family of trajectories defined by the PST is a function of the input variables via the predicted parameter sets. That is, informativity of the input can be directly related to the size of the PST. This feature is now used to define the cost function, using the stage cost $J_\mathcal{S}(\hat{\mathbb{X}}_{l|k},v)$ and terminal cost $J_\mathcal{T}(\hat{\mathbb{X}}_{N|k},k)$ 
\begin{align}\label{eq:CostFunction}
\begin{split}
   & J_\mathcal{S}(\hat{\mathbb{X}}_{l|k},v_{l|k}) =\\
   & \max_{x\in \hat{\mathbb{X}}_{l|k}} \norm{Q(x{-}r_{k{+}l}) }_{\infty} {+} \norm{R( u_{l|k}(x){-}\bar{u}_{l|k})}_{\infty}, \\
    &J_\mathcal{T}(\hat{\mathbb{X}}_{N|k},k) =  \beta(k+N)  \hat{J}_\mathcal{S}(\hat{\mathbb{X}}_{N|k},\bar{u}_{N|k}).
\end{split}
\end{align}
where $\beta(j) = \frac{1-\lambda_c^{T-j}}{1-\lambda_c}$ is a time dependent scaling factor, which reduces the incentive for exploration towards the end of the control task. Overall, the terminal cost is estimating the worst-case cost after the prediction horizon assuming that the reference is fixed and no disturbances act on the system. The cost \eqref{eq:CostFunction} induces performance-based exploration, as exploratory actions are used only when an improvement in performance is predicted.

\subsubsection{Homothetic tubes. }
In the HT formulation, the vertices of the state tube are formulated as linear functions of MPC optimization variables. Thus \eqref{eq:CostFunction} can be simplified to 
\begin{align}\label{eq:CostFunction_HT}
\begin{split}
    &J_\mathcal{S}^H(\hat{\mathbb{X}}_{l|k}^H,v) = \max_{j\in \mathbb{N}_{1}^{q} }\norm{Q(\hat{x}_{l|k}^j{-}r_{k{+}l}) }_{\infty} {+} \norm{R( \hat{u}_{l|k}^j{-}\bar{u}_{l|k})}_{\infty}, \\
\end{split}
\end{align}
which can be written as a linear cost function using an epigraph reformulation  (\cite{boyd2004convex}).

\subsubsection{Flexible tubes. }
In the FT formulation, \eqref{eq:CostFunction} cannot be directly reformulated into a linear cost function in the optimization variables. Hence, the maximization over the state and input terms is separated, to obtain 
\begin{align}\label{eq:CostFunction_FT}
\begin{split}
    J_\mathcal{S}^F(\hat{\mathbb{X}}_{l|k}^F,v) = &\max_{x\in \hat{\mathbb{X}}_{l|k}^F} \norm{Q(x-r_{k+l}) }_{\infty} + \\
   &\max_{x\in \hat{\mathbb{X}}_{l|k}^F} \norm{R( u_{l|k}(x)-\bar{u}_{l|k})}_{\infty},
\end{split}
\end{align}
which can then be reformulated as a linear cost. Note that the \eqref{eq:CostFunction_FT} overestimates the stage cost 
 defined in \eqref{eq:CostFunction}. This can potentially lead to suboptimal exploratory actions, but such approximations are also used in other FT methods in literature (\cite{lu2019robust}) to simplify the resulting online optimization problem.

\section{Dual adaptive MPC}\label{Sec:DualAMPC} \vspace{-2mm}
In this section, HT and FT dual adaptive MPC algorithms are presented using the ingredients described in Sections \ref{Sec:RAMPC} and \ref{Sec:PST_Exploration}.

\subsubsection{Homothetic tubes. }\vspace{-2mm}
In the HT formulation, the optimization variables  are
\begin{align}
\begin{split}
    \gamma_k^H = \Big\{ & \big\{z_{l|k}, \alpha_{l|k}, v_{l|k}, \Bar{u}_{l|k}, \hat{z}_{l|k}, \hat{\alpha}_{l|k}, \{\Lambda^{H,j}_{l|k}\}_{j=1}^{q} \big\}^{N}_{l=0}, \\  
    & \big\{  \{\hat{\Lambda}^{H,j}_{l|k}\}_{j=1}^{q} \big\}^{N-1}_{l=0}, \big\{\hat{\Psi}_{l|k}, \hat{h}_{\theta_{l|k}}\}^{N_p}_{l=1} \Big\},
\end{split}
\end{align}
and the online optimization problem is
\begin{align}\label{eq:HT_OptProb}
\begin{split}
    \min_{\gamma_k^H} \quad  J_\mathcal{T}^H(\hat{\mathbb{X}}_{N|k}^H,k) + \sum_{l-0}^{N-1} J_\mathcal{S}^H(\hat{\mathbb{X}}_{l|k}^H,v_{l|k} )  \\
   \text{ s.t. }  \quad  \eqref{eq:SetpointSubspace}, \eqref{eq:HT_RST}, \eqref{eq:HT_TermCond}, \eqref{eq:xhat}, \eqref{eq:EPPS_RHS_Dual}, \eqref{eq:HT_PST_propagation}.
\end{split}
\end{align}
The dual adaptive MPC algorithm using HT is described in Algorithm \ref{Alg:HT_DAMPC}, where $\hat{u}^*_{k,0}$ is the optimal value of $\hat{u}_{k,0}$.
\begin{algorithm}[h]
\caption{Homothetic dual adaptive MPC}\label{Alg:HT_DAMPC}   
\begin{algorithmic}[1]
    \Statex \textbf{Offline:}  
    \State Choose $\tau, N, N_p$, $\bar{\theta}_0$
    \State Design $K$, $H_x$ satisfying Assumptions \ref{As:Feedback}, \ref{As:Hx}
\setcounter{ALG@line}{0}
    \Statex \textbf{Online:} At each time step $k\ge0$:    
    \State Obtain the measurement $x_k$
    \State Compute $\Theta_k$ and $\bar{\theta}_k$ using set-membership and \eqref{eq:LMS}
    \State Solve \eqref{eq:HT_OptProb}
    \State Apply $ \hat{u}^*_{k,0}$ to the system
\end{algorithmic}
\end{algorithm}

\subsubsection{Flexible tubes. }\vspace{-2mm}
The optimization variables in the FT formulation are
\begin{align}
    \gamma_k^F {=} \Big\{ \big\{h_{x_{l|k}}, \hat{h}_{x_{l|k}}, \Bar{u}_{l|k}, v_{l|k}\big\}^{N}_{l=0}, \big\{\hat{\Psi}_{l|k}, \hat{h}_{\theta_{l|k}}\}^{N_p}_{l=1}\Big\},
\end{align}
and the online optimization problem is
\begin{align}\label{eq:FT_OptProb}
\begin{split}
    \min_{\gamma_k^F} \quad J_\mathcal{T}^F(\hat{\mathbb{X}}_{N|k}^F,k) + \sum_{l-0}^{N-1} J_\mathcal{S}^F(\hat{\mathbb{X}}_{l|k}^F,v_{l|k} )   \\
   \text{ s.t. } \quad  \eqref{eq:SetpointSubspace}, \eqref{eq:FT_RST}, \eqref{eq:FT_TermCond}, \eqref{eq:xhat}, \eqref{eq:EPPS_RHS_Dual}, \eqref{eq:FT_PST_propagation}.
\end{split}
\end{align}
The dual adaptive MPC algorithm using flexible tubes is described in Algorithm \ref{Alg:FT_DAMPC}. 
\begin{algorithm}[t]
\caption{Flexible dual adaptive MPC}\label{Alg:FT_DAMPC}   
\begin{algorithmic}[1]
    \Statex \textbf{Offline:}  
    \State Choose $\tau, N, N_p$, $\bar{\theta}_0$ and $\mu$
    \State Design $K$, $H_x$ satisfying Assumptions \ref{As:Feedback} and \ref{As:Hx}             
    \State Compute $\Lambda_{1,l|k}^{F,i}$, $\Lambda_{2,l|k}^{F,i}$ for $i\in \mathbb{N}_1^{n_x}, l\in \mathbb{N}_0^{N-1}$ from \eqref{eq:Flexible_Constraints_Lambda} and \eqref{eq:Flexible_Propagation_Lambda}
\setcounter{ALG@line}{0}
    \Statex \textbf{Online:} At each time step $k\ge0$:    
    \State Obtain the measurement $x_k$
    \State Compute $\Theta_k$ and $\bar{\theta}_k$ using set-membership and \eqref{eq:LMS}
    \State Solve \eqref{eq:FT_OptProb}
    \State Apply $ \hat{u}^*_{k,0}$ to the system 
\end{algorithmic}
\end{algorithm}

\begin{remark}
    In numerical simulations, it was observed  that \eqref{eq:FT_OptProb} is difficult to solve. One possible reason for this was the recursive computation of  $\hat{\Theta}_{l|k}$ in \eqref{eq:EPPS_RHS_Dual} resulting in a large number of bilinearities. A relaxation which worked well in practice was to compute $\hat{\Theta}_{l|k}$ using $\Theta_k$ and all the predicted measurements in $\{\hat{x}_{i|k}\}_{i=-\tau}^{l}$, resulting in the following approximation of \eqref{eq:EPPS_RHS_Dual}
    \begin{align} \label{eq:EPPS_RHS_Dual_apx}
        \begin{split}
            \hat{h}_{\theta_{l|k}} =  \hat{\Psi}_{l|k} \begin{bmatrix}
               h_{\theta_k}\\ \tilde{h}_{\Delta_{l|k}}
                \end{bmatrix}, \: 
            \:  \hat{\Psi}_{l|k} \begin{bmatrix}
                H_\theta \\ \tilde{H}_{\Delta_{l|k}}
                \end{bmatrix}  = H_{\theta},  \:  \hat{\Psi}_{l|k}  \geq 0,
        \end{split}
    \end{align}
    where $\tilde{h}_{\Delta_{l|k}}$ and $\tilde{H}_{\Delta_{l|k}}$ are defined according to
    \begin{align}\label{eq:Delta_tilde}
    \begin{split}
    &\{\theta \: | \: \hat{x}_{i+1|k} {-} A(\theta) \hat{x}_{i|k} {-} B(\theta) \hat{u}_{i|k} \in \mathbb{W}, \: \forall i \in \mathbb{N}_{-\tau}^{l-1}\}\\
    &= \{\theta \: | \: \tilde{H}_{\Delta_{l|k}} \theta \le \tilde{h}_{\Delta_{l|k}}\}.
    \end{split}
    \end{align}
\end{remark}

\subsection{Properties}\vspace{-3mm}
\begin{theorem}
    Let Assumptions \ref{As:Feedback}, \ref{As:Hx} hold. If the optimization problem \eqref{eq:HT_OptProb} (or problem \eqref{eq:FT_OptProb}) is feasible at time $k=0$, then it remains feasible for all $k>0$. Moreover,  the closed-loop system formed by any dynamics \eqref{eq:Dynamics}-\eqref{eq:DisturbanceBounds} and the MPC controller satisfies \eqref{eq:Constraints} for all $k>0$.
\end{theorem}\vspace{-3mm}
\begin{pf}
The proof for the FT case is given here. The proof for the HT case is similar, and can be found in \cite{diyou2022}. 

Let \eqref{eq:FT_OptProb} be feasible at some time step $k$.  Before \eqref{eq:FT_OptProb} is solved at  $k{+}1$,  $\bar{\theta}_k$ and $\Theta_k$ are updated to $\bar{\theta}_{k{+}1}$ and $\Theta_{k{+}1}$ respectively, and the reference trajectory is shifted by one time step. It will now be shown that a  feasible solution to \eqref{eq:FT_OptProb} can be constructed using the solution at time $k$. 

It must be noted that the modification of $\bar{\theta}_k$ affects only the cost function (through the PST and the input setpoints), and does not affect the feasibility of \eqref{eq:FT_OptProb}. Moreover, $\hat{\mathbb{X}}^F_{l|k} \subseteq \mathbb{X}^F_{l|k}$ and the RST also satisfies the PST dynamics \eqref{eq:FT_PST_propagation}. Therefore, one only needs to find a feasible solution for the RST (defined by the variables $  \{h_{x_{l|k+1}}\}_{l=0}^{N-1}, \{ v_{l|k+1} \}_{l=0}^{N-1} $) and the terminal set (defined by $v_{N|k+1}$ and $ h_{x_{N|k+1}}$).

Consider the variables 
    \begin{align*}\label{eq:RecFeas1}
    \begin{split}
        v_{l-1|k+1} &= v_{l|k}, \: \forall l \in \mathbb{N}_1^{N-1};  \quad  v_{N-1|k+1}=  v_{N|k+1} = v_{N|k} ,
    \end{split}
    \end{align*}
    which ensure that  $u_{l-1|k+1}(x) = u_{l|k}$ for $l\in \mathbb{N}_{1}^{N}$. Because the updated parameter set satisfies $\Theta_{k+1} \subseteq \Theta_k$ from \eqref{eq:SetInclusion}, a shifted RST from time $k$  also satisfies \eqref{eq:FT_RST} at $k+1$.

    Next, the invariance of the terminal set is used to define a feasible $h_{x_{N|k+1}}$, so that the terminal set remains the same. This is achieved by choosing
    \begin{align}
        h_{x_{N|k+1}} = h_{x_{N|k}} + H_x r_{k+N} - H_x r_{k+N+1},
    \end{align}
    which is a feasible solution for the terminal conditions \eqref{eq:FT_TermCond}.

    Therefore, \eqref{eq:FT_OptProb} is feasible at time $k+1$, and by induction, it is recursively feasible. Moreover, the constraints \eqref{eq:Constraints} are satisfied by design, because $\theta^* \in \Theta_k$.
\end{pf}
\vspace{-2mm}
To the best of the authors' knowledge, FT methods have been proposed only for regulation tasks (i.e., reference is the origin) in literature. Thus, in addition to the novelty of the dual control formulation, Algorithm \ref{Alg:FT_DAMPC} provides a novel design methodology for recursively feasible tracking MPC controllers using an FT formulation. 

\section{Numerical example} \vspace{-2mm}
In this section the performance of the proposed algorithms will be compared to each other, and existing methods from the literature. The controller using  Algorithm \ref{Alg:HT_DAMPC} will be denoted as (HT-ESM) and Algorithm \ref{Alg:FT_DAMPC} as (FT-ESM), where ESM stands for exact set-membership. The performance will be compared to the HT algorithm from \cite{parsi2022explicit} denoted as (HT-ASM), because it approximates set membership using set-intersections. In addition a comparison with the FT algorithm from \cite{lu2019robust} is also shown, and denoted as (FT-P) because it performs passive exploration.  Note that \cite{lu2019robust} designs a controller for a regulation task, but the tracking formulation was added to FT-P for a meaningful comparison. The code to simulate the example is available in a public repository (\cite{ESMRepo}).

The model of the system is parameterized by the matrices
\begin{equation}\label{eq:ExampleSystem}
\begin{array}{l l l}
A_0 = \begin{bmatrix} 0.9 &  0.5 \\ 0.2 & 0.8 \end{bmatrix}, & A_1 = \begin{bmatrix} 0.1 &  0 \\ 0 & 0.2 \end{bmatrix}, &A_2 = \begin{bmatrix} 0 &  0 \\ 0 & 0 \end{bmatrix}, \\
B_0 = \begin{bmatrix} 1 & 0.5\\ 0.2& 0.775 \end{bmatrix},  & B_1 = \begin{bmatrix} 0 &  0 \\ 0 & 0 \end{bmatrix},  & B_2 = \begin{bmatrix} 0 & 0.2\\ 0 & 0.35 \end{bmatrix},  \\
\end{array} 
\end{equation}
The uncertainty set is $\Theta = \{\theta\in\mathbb{R}^{2}|\: ||\theta||_\infty \le 1.2 \}$. The constraint set $\mathbb{Z}$ is defined as 
\begin{align*}
	\mathbb{Z} &= \left\{(x,u){\in} \mathbb{R}^{2\times 2}\left|\: \begin{array}{rl}
	||x||_\infty &\le 3, 
	||u||_\infty \le 2  
	\end{array}\right.  \right\},
\end{align*}
and the disturbance set is $\mathbb{W}:= \{w\in\mathbb{R}^2|\: ||w||_\infty \le 0.1\} $. The cost function is defined by $Q = 4I_2$ and $ R = I_2$. 

The parameters for all the controllers are chosen to be $N = 8$, $N_p=5$, $\tau=1$ and $\bar{\theta}_0 = [0.1,0.1]^T$. A piecewise constant reference trajectory with five different setpoints is to be tracked, with a simulation length of $T{=}100$ time steps. The simulations are performed for 500 different realizations of the true parameter $\theta^* \in \Theta$ and disturbance sequences $w_k \in \mathbb{W}$. The parameters and disturbances are chosen randomly from a uniform distribution. All the optimization problems were implemented using YALMIP and solved using MOSEK and IPOPT on an Intel Xeon Gold 5118 processor with 2GB RAM. The FT-P method solves a linear program, whereas all the dual controllers solve nonconvex optimization problems. The average computation times for the online optimization used in  FT-P and FT-ESM methods are $4.14 s$ and $6.77 s$, and that in HT-ASM and HT-ESM methods are $6.05s$ and $10.54s$ respectively. 

\begin{figure}[t]
    \centering
    \includegraphics[scale=1]{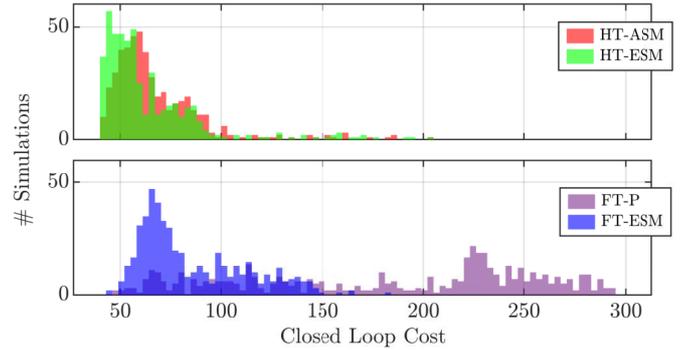}
    \caption{Distribution of closed loop costs over 500 random realizations of $\theta^*$ and $w_k$. } 
    \label{fig:cost}
\end{figure}

The performance of the controllers is compared in Figure \ref{fig:cost}, where the distribution of closed loop costs achieved by each controller is shown. First, it can be seen that the FT-P controller results in high closed loop costs compared to the dual controllers, because the exploration is passive. Such a result has been demonstrated for HT methods in \cite{parsi2022explicit}, and now is replicated for FT methods. Second, the use of exact set-membership reformulation in HT-ESM results in lower costs compared to HT-ASM, because the latter overestimates the information that can be obtained from dual control. Third, it can be seen that both the HT methods perform better compared to the FT-ESM method. This is because the FT method computes Lagrange multipliers offline, uses the approximation in \eqref{eq:EPPS_RHS_Dual_apx} and simplifies the cost function in \eqref{eq:CostFunction_FT} so that the optimization problem is tractable. Thus, although the FT parameterization allows for efficient computation, it results in a loss in performance. 

It must be noted that  the large variation in closed-loop costs for a given controller is because the true parameter affects the reference tracking performance. Moreover, the ESM methods are susceptible to high costs in cases when the initial guess $\bar{\theta}_0$ is far from the true parameter $\theta$. It can be seen that the number of such cases is low, and future work will focus on reducing the dependence on $\bar{\theta}_0$.

\begin{figure*}[!th]
    \centering
    \includegraphics[scale=0.9]{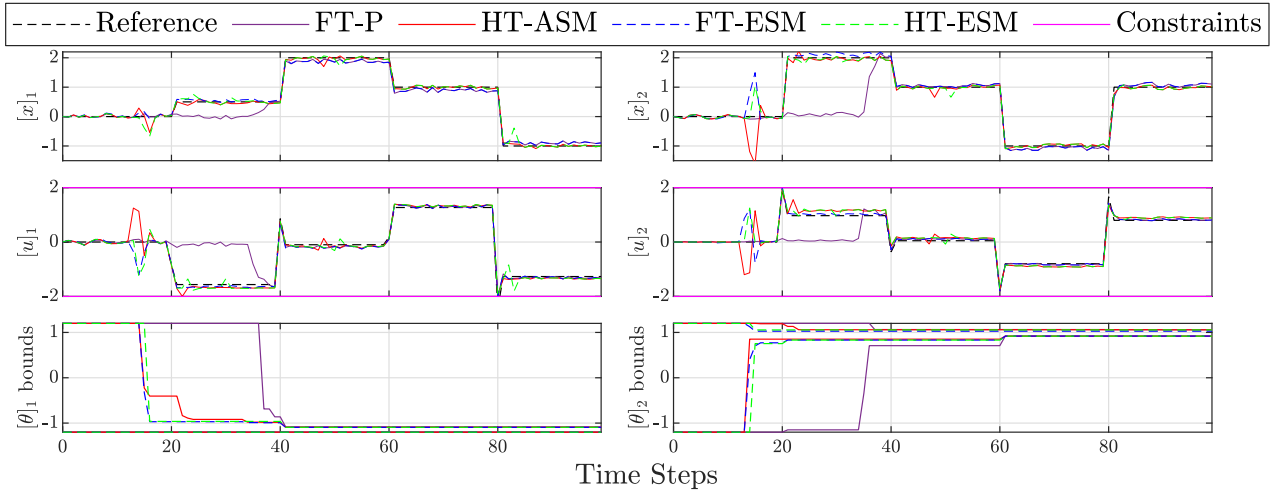}
    \caption{State and input trajectories of the system under four different controllers, along with upper and lower bounds on parameters. The true parameter is $\theta^* = [-1.16,0.96]$.    } 
    \label{fig:tracking}
\end{figure*}

Figure \ref{fig:tracking} shows the state trajectories, input trajectories and the bounds on the parameters for one specific realization of the true parameter and disturbance sequence. It can be seen that that all the controllers ensure constraint satisfaction. The FT-P controller initially does not use the inputs due to the large uncertainty sets and lack of incentive to explore. This results in a large tracking error until passive exploration reduces the uncertainty. In contrast, exploratory actions are used by all the dual controllers, as seen in the deviation of the trajectories from the setpoint between time steps 12 and 20. This improves the overall closed loop costs. The FT-P controller has a cost of 184.8, whereas the cost is 63.0 for FT-ESM, 47.3 for HT-ASM, and 43.5 for HT-ESM. It can also be seen that the HT-ASM controller has worse identification performance compared to HT-ESM (see time steps 10-25), owing to the approximation of set-membership equations using set intersections.

% \begin{figure}[t]
%     \centering
%     \includegraphics[scale=1]{tracking_and_parameter_estimation_6row.eps}
%     \caption{\ani{Tracking and estimation}} 
%     \label{fig:tracking}
% \end{figure}

\section{Conclusion} \vspace{-2mm}
In this work, a performance-based dual control framework is proposed to navigate the dual control trade-off in optimal tracking problems. This framework has been demonstrated with two popular tube MPC formulations used in the adaptive MPC literature. Dual control is achieved by an exact reformulation of set-membership identification within MPC optimization. The proposed method ensures recursive feasibility and robust constraint satisfaction, and numerical simulations demonstrate performance improvement compared to existing methods. Future work involves reducing the dependence of the dual effect on the initial parameter estimate, and extending the method to nonlinear adaptive MPC algorithms.
\bibliography{biblio_ESM}             
                                                   
\end{document}